\documentclass[10pt]{article}
\pdfoutput=1
\usepackage{amssymb,amsmath,mathrsfs,enumerate}
\usepackage{graphicx,rotate,multicol}
\usepackage{wrapfig}
\usepackage[colorlinks=true,
            linkcolor=red,
            urlcolor=blue,
            citecolor=blue]{hyperref}
\usepackage{color}
\usepackage{soul}
\usepackage{cite}
\usepackage{bbold}
\usepackage{verbatim}
\long\def\rpl#1!!#2!!{\textcolor{red}{#1} \textcolor{blue}{#2}}

\DeclareMathOperator{\Tr}{Tr}

\usepackage{cleveref}
\usepackage[makeroom]{cancel}
\usepackage[normalem]{ulem}
\usepackage{amsmath,amssymb}
\usepackage{slashed}
\usepackage{xcolor}
\usepackage{graphicx}
\usepackage{cite}
\usepackage{pdflscape}
\usepackage{multirow}
\usepackage[thinlines]{easytable}
\usepackage{url}
\usepackage[utf8]{inputenc}
\usepackage{longtable}
\usepackage{booktabs}
\usepackage{graphicx,floatflt,amssymb,rotate, cancel}
\usepackage{mathrsfs}
\usepackage{amssymb}
\usepackage{amsmath}
\usepackage{subcaption}
\usepackage{float}
\usepackage{pstricks}
\usepackage[toc,page]{appendix}
\usepackage[mathscr]{euscript}
\usepackage{color}

\def \order(#1){{\cal O} \left(#1 \right)}

\evensidemargin=\oddsidemargin
\def\mET{E_T \hspace{-1.0em}/\;\:}
\def\Eqn#1{Eq.\ (\ref{#1})}

\allowdisplaybreaks

\begin{document}
\begin{flushright}
    IPMU19-0107
\end{flushright}

\begin{center}
	{\Large \bf Charged Higgs searches in the Georgi-Machacek model \\ at the LHC } \\
	\vspace*{1cm} {\sf Nivedita
          Ghosh$^{a,}$\footnote{tpng@iacs.res.in},~Swagata
          Ghosh$^{b,}$\footnote{swgtghsh54@gmail.com},~Ipsita
          Saha$^{c,}$\footnote{ipsita.saha@ipmu.jp}} \\
	\vspace{10pt} {\small \em 
          $^a$School of Physical Sciences, IACS,
         Kolkata, India \\ 
	     $^b$Department of Physics, University of Calcutta, 92
	       Acharya Prafulla Chandra Road, Kolkata 700009, India \\
		$^c$Kavli IPMU (WPI), University of Tokyo, Kashiwa, 277-8583, Japan}
	
	\normalsize
\end{center}

\begin{abstract}
The Georgi-Machacek Model allows for a large triplet vacuum expectation value (vev) with a custodial symmetric scalar potential. 
This manifestly leads to large charged Higgs coupling to SM fermions thus enhancing the detection probability of the model at the 
Large hadron collider.
We show that the latest bound from LHC searches on singly charged Higgs scalar decaying to top and bottom quark restricts  
the triplet vev from above. We demonstrate that the combined limit from theoretical constraints and the latest Higgs data together with
this charged Higgs bound can already curb the model parameter space significantly. Further, we propose the charged Higgs 
decaying to $W^{\pm}$ boson and a SM-like Higgs as a prospective channel to probe the yet unbounded model parameter space at the future run of the LHC.
A simple cut-based analysis for this channel has been made showing the parameter region that can be probed with
$5\sigma$ significance at the high luminosity LHC run.
\end{abstract}

\bigskip

\section{Introduction} \label{sec:intro}
The Standard Model (SM) of particle physics has achieved its triumph after the discovery of a 125 GeV scalar resonance at the Large Hadron Collider (LHC)~\cite{Aad:2012tfa,Chatrchyan:2012xdj} experiment.
Nevertheless, the quest for new physics is yet to accomplish. Possibility of having an exotic particle in the realm of current collider reach still can not be discarded and it may 
just need a closer look at the present LHC data. In the second run of the LHC with increased energy and luminosity, a greater amount of data is already
available for more detailed and intricate analysis. 

Among many, a charged scalar particle is one of the appealing exotic candidates in particle phenomenology and has been
searched for long. 
 Most of the beyond Standard Model (BSM) scenarios render singly charged scalar candidate and it has always remained an important probe for the models with SM scalar sector 
extensions~\cite{Cheung:2002gd, Dev:2013ff, Coleppa:2014cca, Bhattacharyya:2014oka, Han:2015hba,Mitra:2016wpr,Arbey:2017gmh,Cen:2018okf,Guchait:2018nkp,Ghosh:2017pxl,
Abbas:2018pfp,Ghosh:2018jpa,Babu:2019mfe}. 
In this paper, we will talk about probing the Georgi-Machacek (GM)
model~\cite{Georgi:1985nv,Chanowitz:1985ug,Gunion:1989ci}, a variant of scalar triplet extension of the SM, through the searches of a singly-charged scalar at the current and future LHC run. The usual Higgs triplet model (HTM) is favored for the explanation on the neutrino mass generation but 
it suffers from electroweak (EW) $\rho$-parameter constraint. 
The GM model, on the other hand, consists of two scalar triplets, one real and one complex, thus, preserving the custodial SU(2) symmetry and
lifting the bound on the triplet vacuum expectation value (vev). 
This large triplet vev induces a large mixing between the GM and the SM sector leading to interesting search processes for the non standard scalars. 
In particular, the charged scalar of the GM model couples to the SM fermions with a strength directly proportional to the vev $(v_t)$ and a significantly large $v_t$ enhances the detection 
possibility of such charged scalar at the collider. Such decay processes are highly suppressed in the HTM
and can thus be an important probe of the GM model. 

Besides the two CP-even scalars, the lightest of which resembles the SM-like Higgs with mass 125 GeV, 
there lies seven scalar particles in the GM particle spectrum arising from the custodial triplet and fiveplet respectively. 
The singly-charged scalar of the custodial fiveplet ($H_5^\pm$), however, has no tree-level interaction with the SM fermions. 
The singly charged scalar of the custodial triplet ($H_3^\pm$) can have significant tree level coupling to SM fermion-anti-fermion pair
due to the large overlap with the Higgs field corresponding to the triplet vev.
In particular, $H_3^{+}$ may acquire a considerable branching ratio (BR) to the $t\bar b$ channel in the kinematically possible mass region. 
Recently, the ATLAS collaboration~\cite{Aaboud:2018cwk} has reported their latest analysis on charged Higgs searches with the data from the 13 TeV run of the LHC 
with 36 $fb^{-1}$ luminosity
and has provided a model independent bound on the production cross-section times the BR for charged Higgs decaying into $t\bar b$ mode. 
Following their analysis, in this paper, we aim to put constraints on the GM model parameter space in the physical basis in terms of 
the scalar masses and mixing angles. This limit will definitely put an upper bound on $v_t$ for a particular charged Higgs mass corresponding to
its BR. 

On the other hand, the theoretical constraints on the GM scalar potential from EW vacuum stability and perturbative
unitarity at tree-level already limits $v_t$ from the lower end.  Simultaneously, the charged scalar masses are also restricted from above due to
the unitarity constraints. We show the bounds for two variants of GM potential, with and without the trilinear terms where the potential with
vanishing trilinear term results in a more stringent bound. 
Additionally, the latest LHC Higgs data also cause a severe constraint in the model parameter space.
Therefore, a combination of theoretical constraint, Higgs data and the limit from charged Higgs analysis by ATLAS will put 
a definitive restriction on the GM model. In this study, we have given a comprehensive description in this regard.
Furthermore, as a future endeavor, we show how the region allowed by the above
bounds can be probed at the future run of the LHC. For this, we propose the other competent decay channel of $H_3^\pm \to W^\pm h$~\cite{Coleppa:2014cca}, 
$h$ being the 125 GeV Higgs resonance. This decay mode is dominant for large neutral scalar mixing angle $\alpha$ and can be
an effective probe in the region where the ATLAS limit is relaxed. It should be mentioned that though the GM model have been studied
extensively in the post-Higgs discovery era~\cite{Kanemura:2012rs,Chiang:2012cn,Hartling:2014aga,Hartling:2014zca,Chiang:2014bia,
Chiang:2015amq,Chang:2017niy,Degrande:2017naf,Blasi:2017xmc,Biswas:2018jun,Das:2018vkv,Chiang:2018cgb,
Banerjee:2019gmr}, implication of charged Higgs searches in association with the latest Higgs data to explore the model 
parameter space has not been discussed so far.

The paper is organized in the following manner. In sec.~\ref{sec:model}, we introduce the model briefly and explain
the theoretical as well as constraint from Higgs data on the model parameter space. In sec.~\ref{sec:ATLAS}, we explain
the production and decay of the charged Higgs state $H_3^\pm$ for various parameter space and portray the
limits provided by the ATLAS analysis. In sec.~\ref{sec:collider}, we demonstrate our analysis for the future probe of the model
 at the LHC
and, in sec.~\ref{sec:results} we show our findings collectively. Finally, we conclude in sec.~\ref{sec:summary}.

\section{The Model } \label{sec:model} 
Here, we present a brief description of the scalar potential of the GM model. 
in addition to the SM particle content, the scalar sector of the 
GM model~\cite{Georgi:1985nv,Chanowitz:1985ug,Gunion:1989ci} consists of one real $SU(2)_L$ triplet $\xi$ 
and one complex $SU(2)_L$ triplet $\chi$ with hypercharges  $Y=0$  and $Y=2$ forming a bi-triplet $X$.  
Now, the most general scalar potential for the GM model can be written as~\cite{Hartling:2014zca,Das:2018vkv}

\begin{eqnarray}
V(\Phi,X) &=& \frac{\mu_{\phi}^2}{2} \Tr(\Phi^\dagger \Phi) + \frac{\mu_{X}^2}{2} \Tr(X^\dagger X) + \lambda_1[\Tr(\Phi^\dagger \Phi)]^2 +
\lambda_2 \Tr(\Phi^\dagger \Phi)\Tr(X ^\dagger X) \nonumber \\ 
&& + \lambda_3 \Tr(X^\dagger X X^\dagger X) + \lambda_4 [\Tr(X^\dagger X)]^2 - \lambda_5 \Tr(\Phi^\dagger \tau_a \Phi \tau_b)\Tr(X^\dagger t_a X t_b) \nonumber \\
&& - M_1\Tr(\Phi^\dagger \tau_a \Phi \tau_b)\left( U X U^\dagger \right)_{ab} - M_2\Tr(X^\dagger t_a X t_b)\left( U X U^\dagger \right)_{ab} \,,
\label{e:potential}
\end{eqnarray}
where, $\tau_a \equiv \sigma_a/2$, ($a=1,2,3$) with $\sigma_a$'s  being 
the Pauli matrices and  $t_a$'s are the generators of triplet representation of
$SU(2)$ which are expressed as
\begin{eqnarray}
t_1 = \frac{1}{\sqrt{2}}\left(\begin{array}{ccc}
0 & 1 & 0 \\
1 & 0 & 1 \\
0 & 1 & 0 \\
\end{array}\right)\,, \qquad
t_2 = \frac{1}{\sqrt{2}}\left(\begin{array}{ccc}
0 & -i & 0 \\
i & 0 & -i \\
0 & i & 0 \\
\end{array}\right)\,, \qquad
t_3 =\left(\begin{array}{ccc}
1 & 0 & 0 \\
0 & 0 & 0 \\
0 & 0 & -1 \\
\end{array}\right)\,.
\label{su2_tripgen}
\end{eqnarray}
Furthermore, the matrix $U$ in the trilinear terms of \Eqn{e:potential} is given by,
\begin{eqnarray}
U = \frac{1}{\sqrt{2}} \left(\begin{array}{ccc}
-1 & 0 & 1 \\
-i & 0 & -i \\
0 & \sqrt{2} & 0 \\
\end{array}\right) \,.
\label{matU}
\end{eqnarray}

It is worth mentioning here that the absence of these trilinear term would enhance the 
symmetry of the potential. Moreover, it should be noted that the GM potential has 
no explicit CP violation since there are no non-Hermitian terms.

Now, the bi-doublet $\Phi$ and the bi-triplet $X$ are represented as
\begin{eqnarray}
\label{e:fields}
\Phi = \left(\begin{array}{cc}
\phi^{0*} & \phi^+ \\
-\phi^- & \phi^0 \\ 
\end{array}\right) \,, \qquad
X= \left(\begin{array}{ccc}
\chi^{0*} & \xi^+ & \chi^{++} \\
-\chi^- & \xi^0 & \chi^+ \\
\chi^{--} & -\xi^- & \chi^0 \\
\end{array}\right) \,.
\end{eqnarray}
After the electroweak symmetry breaking (EWSB), the neutral components of the bi-doublet and the bi-triplet can be expanded around their vacuum expectation values(vev) as
$\phi^0 = \frac{1}{\sqrt{2}} (v_d + h_d + i \eta_d)$, $\xi^0 =  (v_t + h_\xi )$ and $\chi^0 =  (v_t + \frac{h_\chi + i \eta_\chi}{\sqrt{2}})$.
 The equality in vevs to the real and the complex triplets
corresponds to the preserved custodial symmetry of the potential. The EW vev 
in terms of the vevs of the scalar multiplets turns out to be
\begin{eqnarray}
\sqrt{v_d^2 + 8 v_t^2} = v = 246~{\rm GeV} \,.
\label{e:custodial}
 \end{eqnarray}
Thus, there will be two independent minimization conditions corresponding to the two vevs of the
bi-doublet and the bi-triplet ($v_d$ and $v_t$) which can be used to extract the bilinear
coefficients of the potential $\mu_{\phi}^2$ and $\mu_X^2$ as
\begin{subequations}
	\begin{eqnarray}
	\mu_{\phi}^2 &=&  - 4 \lambda_1 v_d^2 - 3\left(2\lambda_2 -\lambda_5\right) v_t^2 + \frac{3}{2} M_1 v_t \,, \\
	\mu_{X}^2 &=&   -\left(2\lambda_2 -\lambda_5\right) v_d^2 - 4 \left(
	\lambda_3+ 3\lambda_4 \right)v_t^2 + \frac{M_1 v_d^2}{4v_t} + 6 M_2 v_t \,.
	\end{eqnarray}
	\label{e:bilinears}
\end{subequations}

The scalar potential consists of a custodial quintuplet $(H_5^{++},H_5^+,H_5^0, H_5^-, H_5^{--})$ 
of common mass $m_5$ and a custodial triplet $(H_3^+,H_0, H_3^-)$ of mass $m_3$. Alongside,
there are two CP-even scalars that are custodial singlets, namely $h$ and $H$ with masses 
$m_h$ and $m_H$ respectively. The neutral scalar mixing angle $\alpha$ diagonalizes 
the CP even sector to obtain the mass eigenstates $h$ and $H$. Hereby, we refrain ourselves from giving a detailed description
of the diagonalization procedure and we refer the reader to Refs.~\cite{Chiang:2012cn,Hartling:2014zca}.
We only show the mass eigenstate for the charged scalars and the custodial singlets.

\begin{subequations}
\begin{eqnarray}
\label{e:ch}
H_5^\pm &=& \frac{1}{\sqrt{2}} \left(\chi^\pm - \xi^\pm \right) \,, \qquad
H_3^\pm = -\sin\beta~ \phi^\pm + \cos \beta \frac{1}{\sqrt{2}} \left(\chi^\pm + \xi^\pm \right) \,, \\
\label{e:alpha}
h &=& \cos\alpha~ h_d + \sin\alpha \left(\sqrt{\frac{1}{3}}h_\xi + \sqrt{\frac{2}{3}}h_\chi \right) \,, \\
H &=& -\sin\alpha ~h_d + \cos\alpha \left(\sqrt{\frac{1}{3}}h_\xi + \sqrt{\frac{2}{3}}h_\chi \right) \,,
\end{eqnarray}
\end{subequations}
where, $h_d,h_\chi,h_\xi$ are the fields corresponding to the neutral components of the doublet and the triplets
after the electroweak symmetry breaking. The angle $\alpha$ represents the neutral mixing angle while $\tan\beta$ is defined as
\begin{eqnarray}
\tan\beta = \frac{2\sqrt{2} v_t}{v_d} \,.
\end{eqnarray}

In total, there are nine independent parameters in the GM scalar potential with 
two bilinears ($\mu_{\phi}^2$ and $\mu_X^2$), five quartic couplings ($\lambda_i$,
$i=1,\dots, 5$) and two trilinear couplings ($M_1$ and $M_2$). Among these, 
the bilinears have already been traded in favor of the vevs, $v_d$ and $v_t$
as in \Eqn{e:bilinears}. Besides the trilinear terms ($M_1,M_2$), the five quartic couplings can now be 
exchanged with the four physical scalar masses, 
$m_5$, $m_3$, $m_H$ and $m_h$ and the mixing angle, $\alpha$.
It should be mentioned here that, $h$ is the lightest among all the $CP$-even scalars
corresponding to the Higgs-like scalar discovered at the LHC with mass $m_h \approx 125$~GeV.
Below, we present the relation between the $\lambda_i$-s with the physical
masses and mixings~\cite{Das:2018vkv}.
\begin{subequations}
	\begin{eqnarray}
	\lambda_1 &=& \frac{1}{8 v^2 \cos^2 \beta}\left(m_h^2 \cos^2 \alpha +m_H^2 \sin^2 \alpha\right) \,, \label{lambda1} \\
		\lambda_2 &=&\frac{1}{12 v^2 \cos \beta \sin \beta}\left( \sqrt{6}\left(m_h^2  - m_H^2\right) \sin 2 \alpha + 12 m_3^2 \sin\beta \cos\beta - 3 \sqrt{2}v \cos\beta M_1 \right) \,, \label{lambda2}\\	
\lambda_3 &=& \frac{1}{v^2 \sin^2 \beta}\left( m_5^2 - 3 m_3^2 \cos^2\beta + \sqrt{2}v \cos\beta \cot \beta M_1 - 3 \sqrt{2} v \sin\beta M_2  \right)  \,, \label{lambda3} \\
\lambda_4 &=& \frac{1}{6 v^2 \sin^2 \beta}\Big( 2 m_H^2 \cos^2 \alpha + 2 m_h^2 \sin^2\alpha - 2 m_5^2 
+6 \cos^2\beta m_3^2  - 3\sqrt{2} v \cos\beta \cot \beta M_1 \nonumber \\
&&    + 9 \sqrt{2} v \sin \beta  M_2 \Big) \,, \label{lambda4} \\
	\lambda_5 &=&  \frac{2 m_3^2}{v^2} -\frac{ \sqrt{2} M_1}{ v \sin \beta} \,. \label{lambda5}
		\end{eqnarray}
		\label{e:masstolam}
\end{subequations}

\subsection{Theoretical Constraint and Higgs data}\label{subsec:theocon}
In this section, we discuss the consequences of theoretical 
constraints, such as the electroweak stability and the perturbative 
unitarity on the GM scalar potential as well as the parameter constraints coming from the 
latest Higgs data. We will thus use \Eqn{e:masstolam} to
translate the scalar couplings in terms of the physical masses and mixings. 
\begin{figure}[ht!]
	\centering
	\includegraphics[width=8cm,height=6cm]{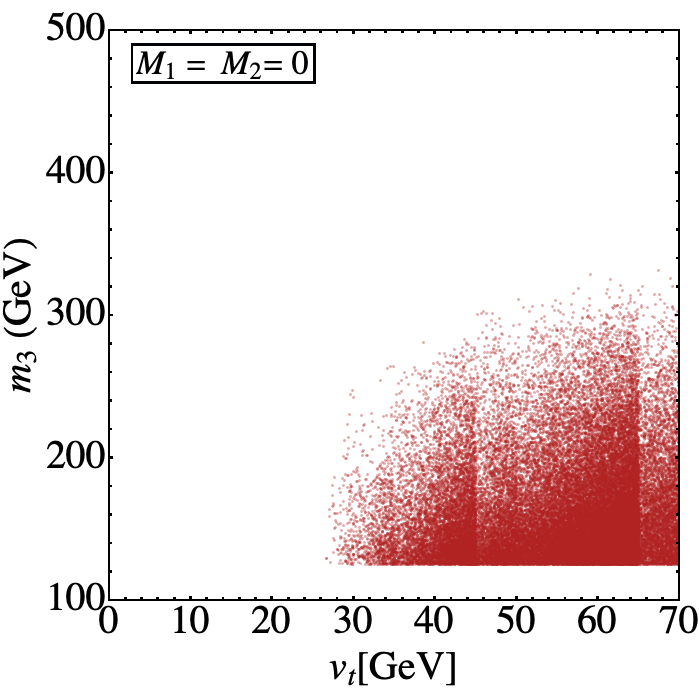}  
	\includegraphics[width=8cm,height=6cm]{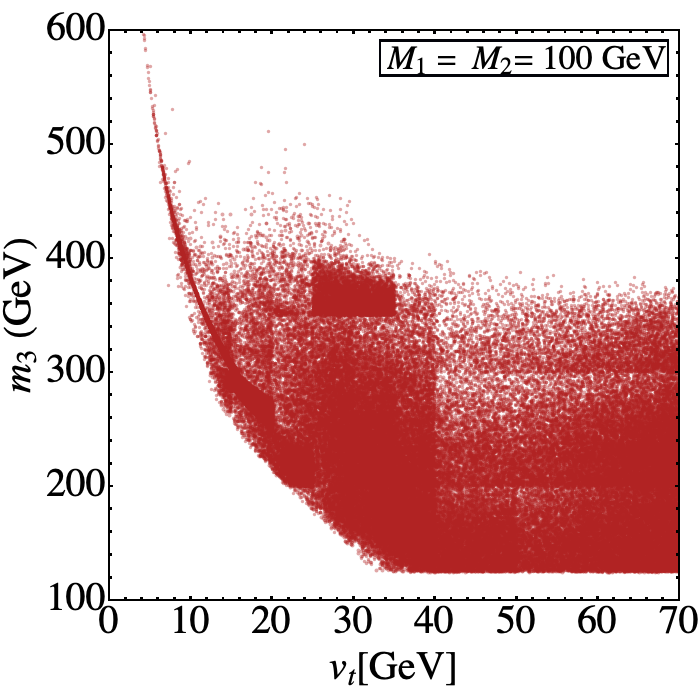} \\
	\includegraphics[width=8cm,height=6cm]{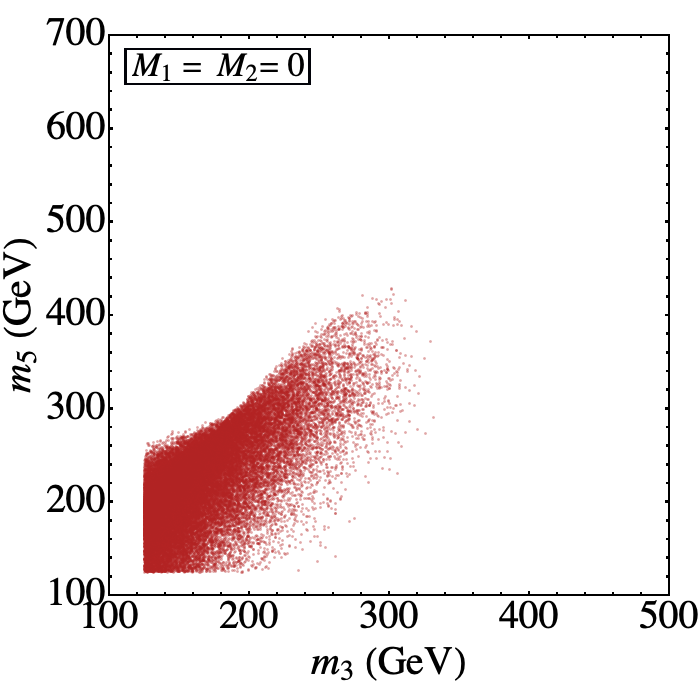} 
	\includegraphics[width=8cm,height=6cm]{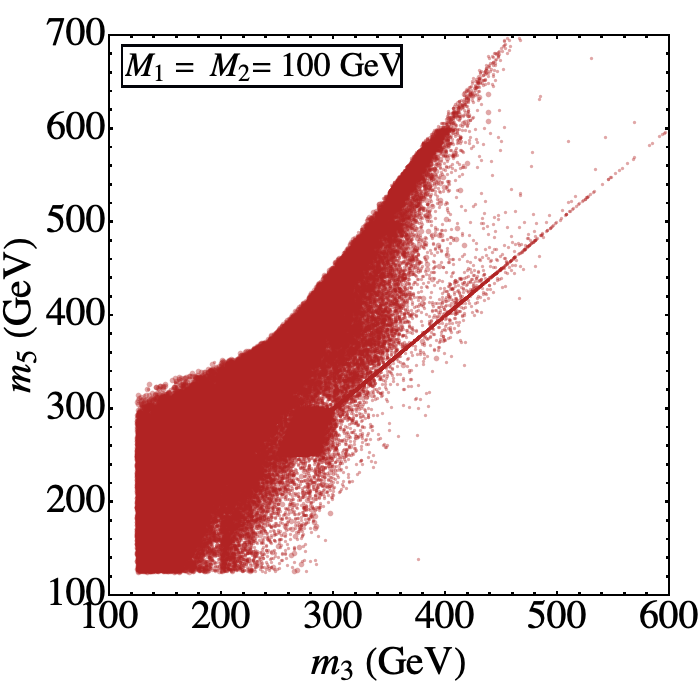} 
	\caption{\em Allowed regions in the $v_t$-$m_3$ plane (upper panel) and in the $m_3$-$m_5$ plane (lower panel) from
		theoretical constraints for zero (left panel) and non-zero (right panel) trilinear couplings.}
	\label{f:theo}
\end{figure}

We first consider the theoretical constraints on the model parameter space. In Fig.~\ref{f:theo}, we show the region allowed simultaneously by the electroweak vacuum stability and 
the perturbative unitarity bounds on the scalar quartic couplings~\cite{Aoki:2007ah,Hartling:2014zca}. For illustration purpose, we
have shown the allowed region in the $v_t$-$m_3$ plane  and in the $m_3$-$m_5$ plane. In doing so, we choose two different
possibilities for the GM scalar potential, with and without the trilinear couplings. For nonzero trilinear couplings, we choose
$M_1=M_2=100~\rm GeV$ as a benchmark. 
As is evident from the plots that the theoretical constraints 
impose a lower bound on the triplet vev $v_t$ as well as an upper bound on the mass of the custodial triplet $m_3$ and custodial fiveplet $m_5$
when the scalar potential does not contain any trilinear terms, i.e. $M_1=M_2=0$. It also draws
a correlation between $m_3$ and $m_5$ as can be seen from the lower panel of  Fig.~\ref{f:theo}.
These limits are actually a direct consequence of the unitarity condition on the scalar quartics as has been discussed in detail in Ref~\cite{Das:2018vkv}.
Briefly, the triplet is required to obtain a reasonable vev $v_t \gtrsim 27 \rm GeV$ to get a simultaneous solution of all the unitarity conditions. 
The limits can however be lifted by the introduction of the non-zero trilinear terms in the potential as can be viewed from the right panel of Fig.~\ref{f:theo}. 
The reason is the additional contribution of the nonzero trilinear couplings to the scalar quartic couplings, given in Eq.~(\ref{e:masstolam}), 
that helped to alleviate the bounds on the masses and the vev $v_t$. Moreover,
with non-zero trilinear terms, the limit for small or vanishing $v_t$ can be achieved for large non standard scalar masses $m_3 \simeq m_5 \simeq m_H \gg m_h$ and
for $M_1 \simeq {\cal O}(v_t)$, this limit is often referred to as the {\it decoupling limit} \cite{Hartling:2014zca}.
\begin{figure}[ht]
	\centering
	\includegraphics[width=6cm,height=6cm]{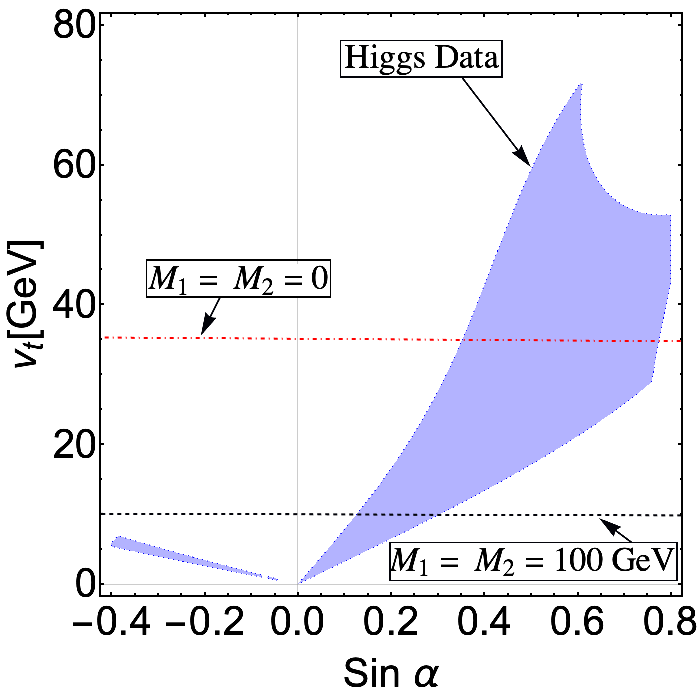}  
	\caption{\em Model parameter space allowed in the ($\sin\alpha$-$v_t$) plane from the latest Higgs data (shaded blue)~\cite{CMS:2018lkl,ATLAS:2018doi} for a benchmark scenario $m_3(m_5)=300(400)$ {\rm GeV}
		allowed by the theoretical constraints. The dot-dashed (red) and dashed (black) lines correspond to the bound coming from theoretical constraint for vanishing and non vanishing trilinear term
		of the GM potential respectively.}
	\label{f:hdata}
\end{figure}	

Next, one should also consider the bounds on the model parameter space placed by the updated 13 TeV LHC Higgs data~\cite{CMS:2018lkl,ATLAS:2018doi}.
Here, it is to be mentioned that the lightest among all the CP-even states $(m_h)$ corresponds to the 125 GeV SM-like Higgs
in our model scenario. Therefore, its tree-level couplings to the SM fermions and vector bosons must be matched with the
LHC data. Additionally, the charged scalar particles coming from both the custodial multiplets of masses $m_3$ and $m_5$ 
contributes to the loop-induced Higgs to diphoton decay mode. Therefore, the observed limit from $h\to\gamma\gamma$ 
decay~\cite{CMS:2018lkl,ATLAS:2018doi} should also be considered for lighter charged Higgs masses. In fact, the combined limits from Higgs decay to
fermions, weak gauge bosons together with di-photon channel severely constraint the model parameter space as shown in Fig.~\ref{f:hdata}.
To implement the current limit, we define the Higgs coupling modifiers as the
ratio of the theoretical model prediction to the SM value, i.e,
\begin{eqnarray}
\kappa_i = \frac{g_{{hii}(GM)}}{g_{{hii}(SM)}}, \quad i=f,W^{\pm},Z,\gamma \,.
\end{eqnarray}
The model predictions for the coupling modifiers are well-known~\cite{Das:2018vkv} and it depend on the mixing angles
$\alpha$ and $\beta$ for a fixed $m_3$ and $m_5$. 
 In Fig.~\ref{f:hdata}, we  show the allowed region in the
$(\sin \alpha-v_t)$ plane from the latest 13 TeV LHC Higgs result~\cite{CMS:2018lkl,ATLAS:2018doi}. The blue shaded region in the figure is allowed by all
Higgs channels including the loop induced diphoton mode. For the diphoton decay mode, we have fixed the 
charged scalar  masses at $m_3=300~\rm GeV$ and $m_5=400~\rm GeV$. The benchmark of such mass points are followed
by the maximum allowed value by the theoretical constraints, as is evident from Fig.~\ref{f:theo}.  Now, in Fig.~\ref{f:hdata}, the solid(red) and dashed(black) lines 
further corresponds to the equivalent theoretical limits on $v_t > 35 (10)~\rm GeV$ for the two cases, $M_1=M_2=0 (100~\rm GeV)$ respectively with $m_3=300~\rm GeV$ following the upper panel of Fig.\ref{f:theo}. Obviously, the limit is more relaxed in the non vanishing trilinear case, allowing $v_t$ 
as low as around 10 GeV. The most important point in Fig.~\ref{f:hdata}, is however, the region allowed 
by the Higgs data. It shows a clear preference towards larger $\sin\alpha$ at large $v_t$ allowing only $\sin\alpha > 0.3$ 
for $v_t \gtrsim 30~\rm GeV$. This has notable consequences on the charged Higgs decay modes as we will explain in our subsequent analysis.
\section{Charged Higgs production and decay}\label{sec:ATLAS}
\begin{figure}[ht!]
	\centering
	\includegraphics[width=8cm,height=6cm]{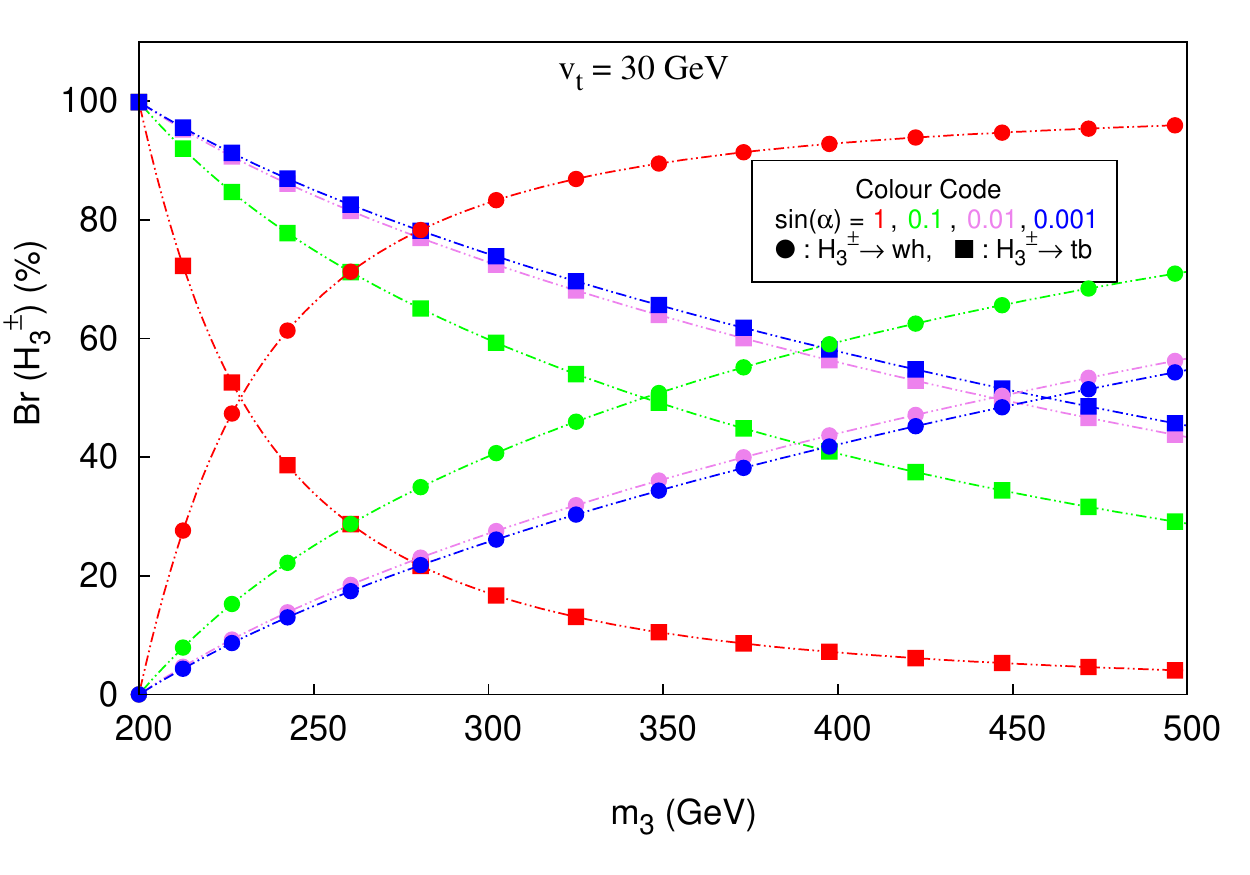}  ~
	\includegraphics[width=8cm,height=6cm]{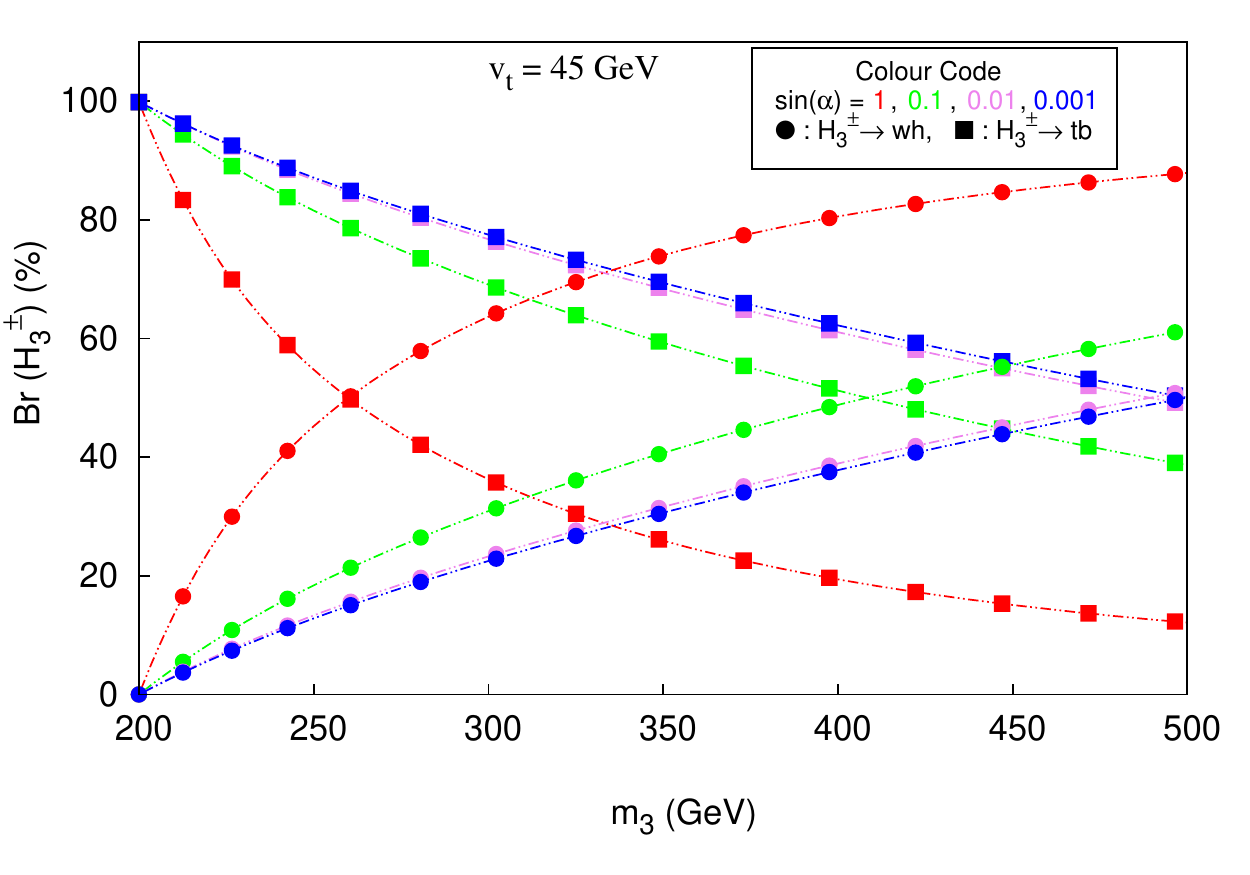} 
	\caption{\em The branching ratio of charged Higgs $H_3^\pm$ for two different choices of the triplet vev, $v_t = 30$ GeV (left panel) and $v_t = 45$ GeV (right panel). 
	The box and circular points define the two distinct decay modes of the charged scalar, $H_3^\pm \to tb$ and $H_3^\pm \to W^{\pm}h$ channel respectively while the
four different colors (red, green, magenta, blue) corresponds to four different orders of magnitude for $\sin\alpha$ (1,0.1,0.01,0.001).   }
	\label{f:h3br}
\end{figure}

\begin{figure}[ht!]
	\centering
	\includegraphics[width=8cm,height=6cm]{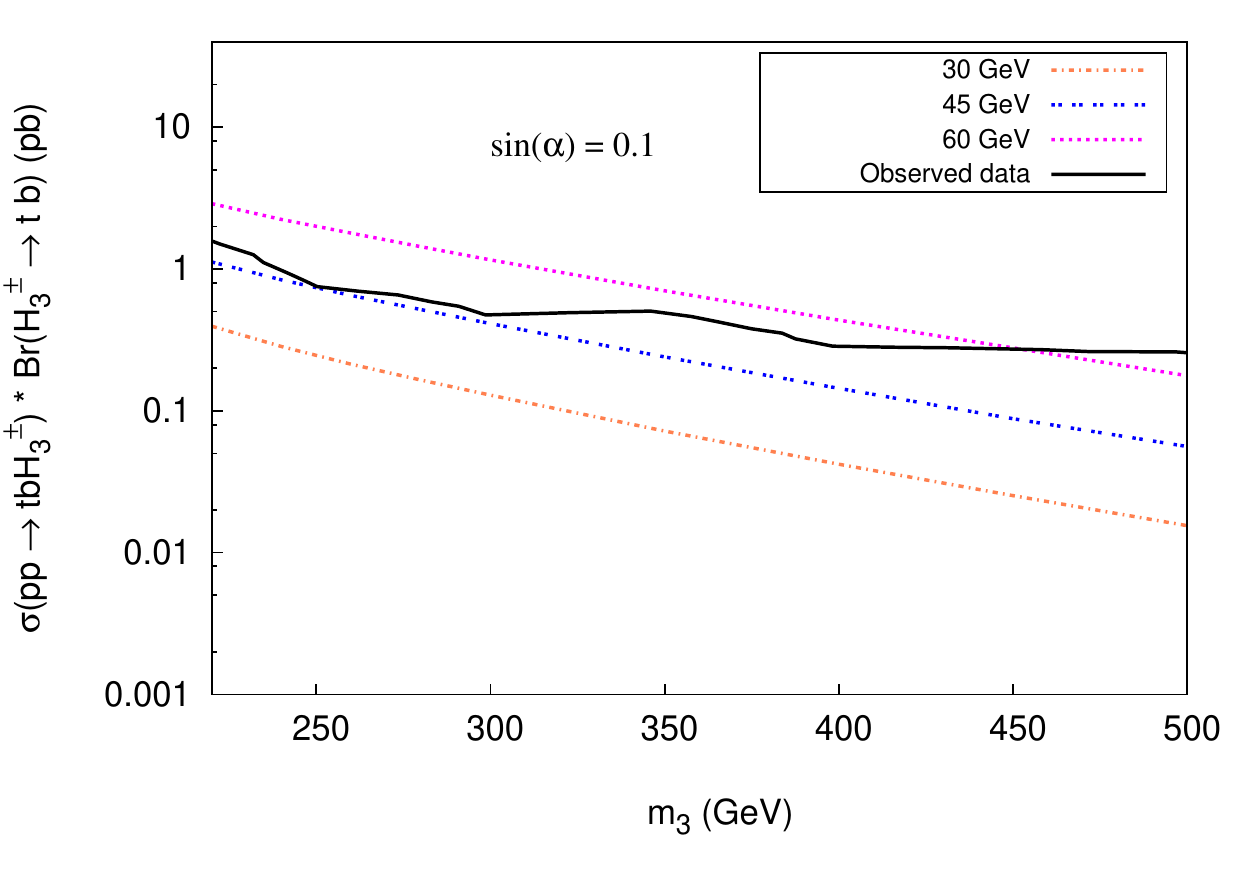}  ~
	\includegraphics[width=8cm,height=6cm]{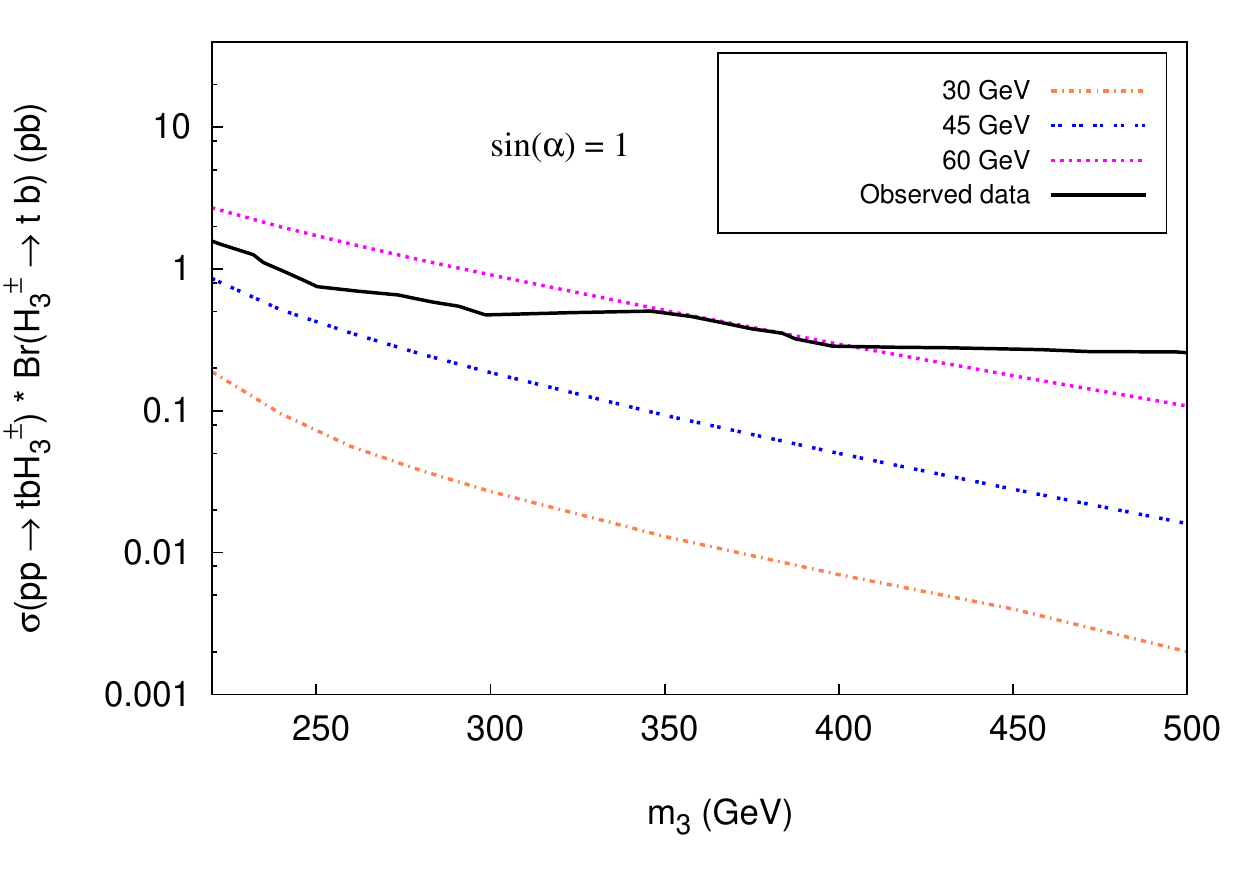} 
	\caption{\em The production cross-section times the branching ratio for the charged Higgs decaying to $tb$ channel is shown for two different values of $\sin\alpha$, 
		0.1 (left panel) and 1.0 (right panel). The three colored lines, orange(dot-dashed), blue(dotted) and magenta(dashed) stands for three different values of triplet vev $v_t = 30,~45~ \&~ 60 $ GeV
		respectively. The solid black line corresponds to the current limit from ATLAS analysis~\cite{Aaboud:2018cwk}.}
	\label{f:h3tb}
\end{figure}

As mentioned earlier, the GM model contains two pairs of singly charged Higsses, one from each of the custodial triplet $(H_3^\pm)$ and the custodial 
quintuplet $(H_5^\pm)$. However, the quintuplet component $H_5^\pm$ does not couple to SM quarks since it has a pure triplet origin, as can also be seen
from Eq.~(\ref{e:ch}). On the other hand, the triplet candidate $H_3^\pm$ can mix with SM quarks through doublet mixing. Therefore, in this article,
we will venture the possibility of limiting or probing the GM model from direct searches of the triplet charged candidate, namely, $H_3^\pm$. 
To begin with, we show the decay branching ratio of this particle for various mass regions. In Fig.~\ref{f:h3br}, we depict the BR of $(H_3^\pm)$ of mass
$m_3$ for two different choices of $v_t = (30,45)~\rm GeV$ and each for four different values of $\sin\alpha = (0.001,0.01,0.1,1.0))$, represented by four different  
color variations. It must be mentioned that our choice of $v_t$ relies on the parameter space available after satisfying the theoretical constraints, as described in Sec.~\ref{subsec:theocon}. 
Although, the non-zero trilinear scenario allows much lower $v_t$,
the BRs has no dependence on
the parameter values $M_1$ and $M_2$. Therefore, in our following analysis, we will mainly focus on $v_t \gtrsim 30~\rm GeV$.
It should however, be mentioned that we have not imposed the Higgs data while plotting Figs. \ref{f:h3br}, \ref{f:h3tb} and \ref{f:h3wh}.

Let us first briefly understand the nature of Fig.~\ref{f:h3br}. As one can see, $H_3^\pm$ has two dominant decay channels, $H_3^+ \to t \bar{b}$ and $H_3^+ \to W^+ h$ (similarly for the anti-particle), $h$  being the 125 GeV resonance.
 It is clear from Fig.~\ref{f:h3br}, that as the mass of 
$H_3$ reaches 205 GeV (threshold to produce $W^{\pm} h$),  the $BR( H_3^{\pm} \rightarrow W^{\pm} h)$ increases considerably with an increase in $\sin\alpha$ 
reducing the $BR(H_3^{\pm} \rightarrow t b)$. This turns out to be the crucial point for the charged Higgs searches in this model. As we showed in Sec.~\ref{subsec:theocon} that the current Higgs data preferred larger $\sin\alpha$ for larger $v_t$. Hence, the combined limit of
theoretical constraints with the Higgs data points towards  an inclination of a lighter charged Higgs mass $m_3 < 300~\rm GeV$ and a larger mixing angle in the CP-even sector. From this viewpoint and referring to Fig.~\ref{f:h3br}, one can interpret that probing the charged Higgs candidate in this model would be more probable in $W^\pm h$ channel instead of the quark final state. 

Nevertheless, it is mandatory to validate the model parameter space first with the current limit on charged Higgs decaying to $tb$ final state, as reported by the ATLAS collaboration~\cite{Aaboud:2018cwk} with the 13 TeV data. There is a similar analysis by CMS collaboration lately~\cite{CMS:2019yat}, however there limits for low charged Higgs mass $m_3 < 500 ~\rm GeV$ are weaker and hence we will only consider the ATLAS bound. 
In Fig.~\ref{f:h3tb}, we draw the limits on the charged Higgs production cross sec times its BR to $tb$ channel, as observed in experiment (further details on cross section measurement are given in section \ref{sec:collider}). The observed limit on the effective cross section from ATLAS for a range of charged Higgs masses has been shown for three distinct triplet vev $v_t = 30,~45~ \&~ 60~\rm GeV$ with two different $\sin\alpha \equiv (0.1,1)$ values. As anticipated, even for low charged Higgs masses 
$m_3 \lesssim 300 ~\rm GeV$, the null result from observed data can only exclude $v_t > 45~\rm GeV$ for $\sin\alpha = 0.1$. The limit is much relaxed for higher mixing angles.

Following the above findings, in the next section we will carry out a simple signal-to-background (cut-based) analysis to predict the efficiency of probing the model parameter space still allowed by theoretical constraints and the present experimental observation made for both the Higgs and the charged Higgs.  
\section{Analysis}\label{sec:collider}
In this section, we address the question that we posed in the last section, i.e. how efficiently we can probe the parameter space of GM model allowed by the experimental and 
theoretical bounds together. Charged Higgs decaying into $W^{+}h$ channel is the signal that we will be interested in. The single production of the charged Higgs can however
be made either in the 4-flavor or in the 5-flavor scheme, where the production processes will be $p p \to t \bar b H_3^+$ and $p p \to t  H_3^+$ respectively.
In our subsequent analysis, we will work in the 5-flavor scheme.
\begin{figure}[ht!]
	\centering
	\includegraphics[width=8cm,height=6cm]{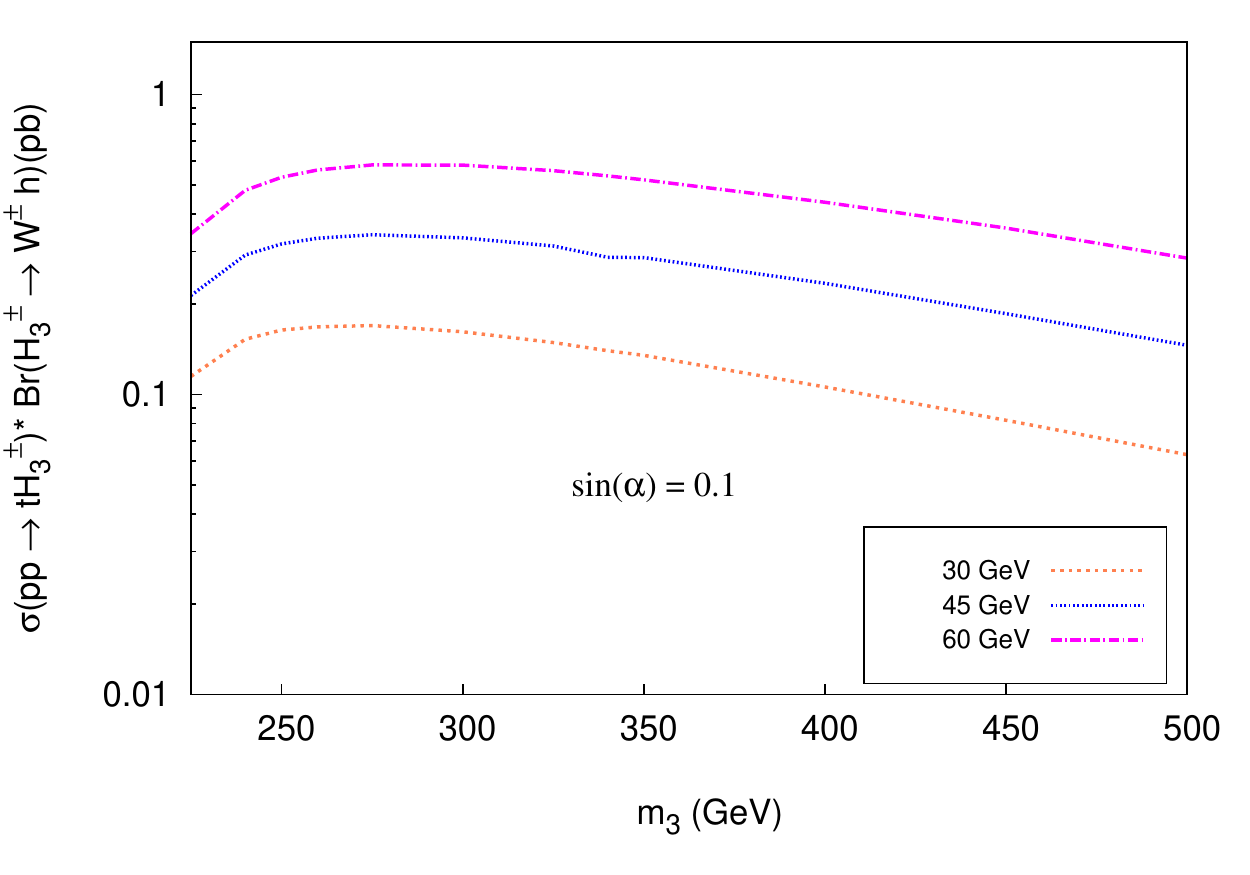}  ~
	\includegraphics[width=8cm,height=6cm]{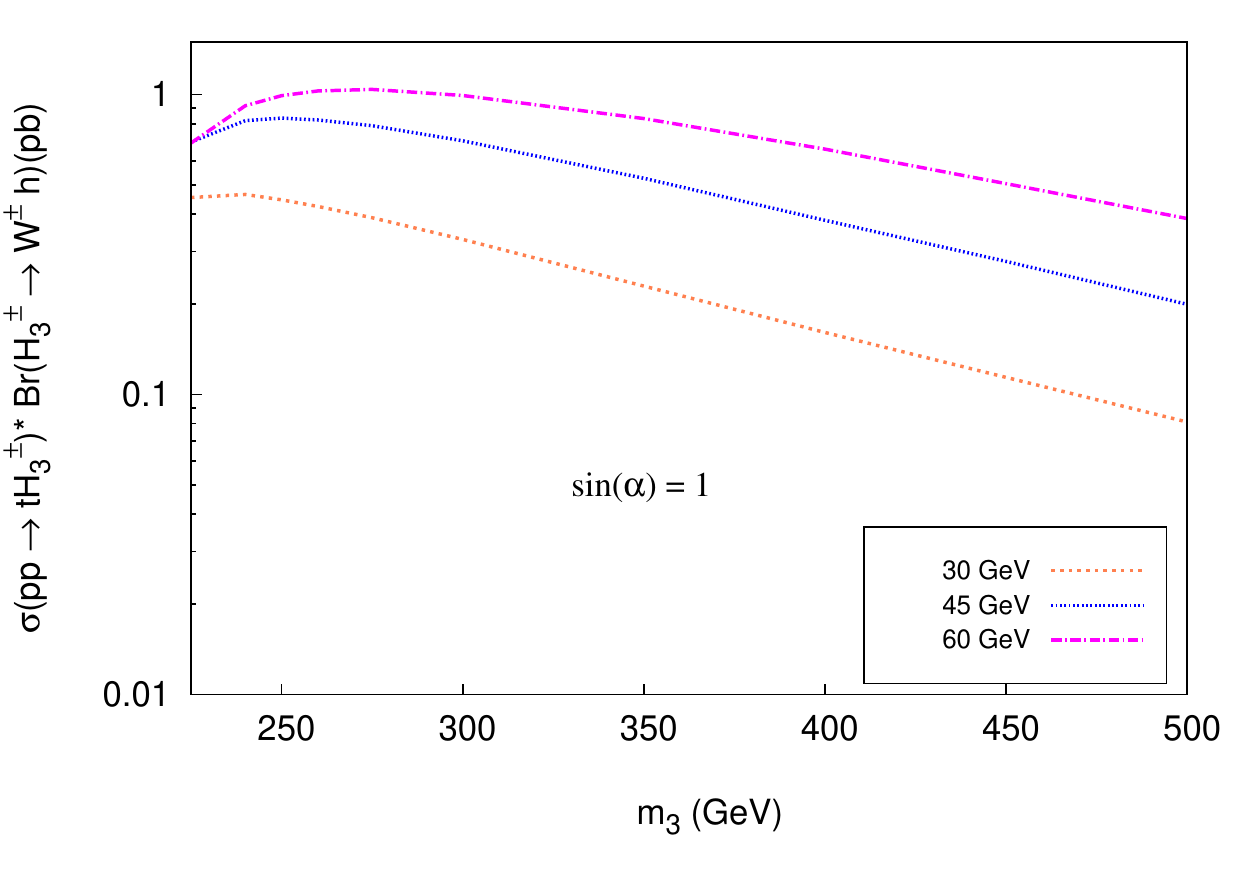} 
	\caption{\em The production cross section times branching ratio plot for charged Higgs decaying to $W^\pm h$ channel for two values of $\sin\alpha =0.1$ (left panel) 
	and $\sin\alpha =1.0$ (right panel). The color coding is same as in Fig.~\ref{f:h3tb}.}
	\label{f:h3wh}
\end{figure}	
In Fig.~\ref{f:h3wh}, we show the production cross section times the BR for $\sin\alpha=1, 0.1$.
Following our previous line of argument, a direct comparison between Fig.~\ref{f:h3tb} and Fig.~\ref{f:h3wh} shows
that for larger $\sin\alpha$, the $W^{+}h$ decay mode for the charged higgs imparts larger effective cross-section.
Hence, we may expect to discover or restrict the model in its full parameter region with larger efficiency 
in the future run of the LHC probing the specified channel. In the following, we will give a simple cut-based 
analysis to give a prediction of probing the GM model at the future LHC run from the charged Higgs searches through
$W^{+}h$ final state.
%
Therefore, we consider the following signal channel:
\begin{eqnarray}\label{e:process}
p p \to t H^-_3 \to b W^+ W^- h \to 3b + 2\ell + \mET \,
\end{eqnarray}
where, we assume that the two $W$ bosons decays leptonically and the Higgs decay yields two b-tagged jets.
Thus as a final state we look for 3 b-tagged jets, two opposite signed leptons and missing transverse energy.
The irreducible SM background contribution will come from $t\bar t j$ with at least one extra hard jet, $t\bar t h$ and $t \bar t V, V={W^{\pm},Z}$ processes. 
It should be noted here that the $t\bar t $ plus jets background has the largest cross section among all and in turn it is the most dominant
background in this analysis. 
Below we present our cut-based analysis for the signal-to-background event ratio computation. 
For our analysis, we choose two benchmark points in consistent with all the above limits, and the relevant parameter values are shown in Table.~\ref{tab:bp}. 
In passing, we would like to mention that our benchmark value for $m_5$ is consistent with the latest bound from ATLAS collaboration~\cite{Aaboud:2018qcu} 
where a doubly-charged Higgs mass between 200 and 220 GeV is excluded at 95\% CL from same-sign $W$ boson signal search.
%
%

The model has been implemented in {\tt FeynRules}~\cite{Alloul:2013bka}
to obtain the UFO model files required for event generation in madgraph. 
We generate both the signal and SM backgrounds events at the Leading Order (LO) 
in {\tt Madgraph5(v2.3.3)}~\cite{Alwall:2014hca} using the {\tt NNPDF3.0} parton distributions~\cite{Ball:2014uwa}. 
The parton showering and hadronisation is done using the built-in {\tt Pythia}
\cite{Sjostrand:2006za} in the Madgraph. 
The showered events are then passed through {\tt Delphes}(v3)~\cite{deFavereau:2013fsa} 
for the detector simulation where the jets are constructed 
using the anti-$K_{T}$ jet algorithm with minimum jet formation radius $\Delta R = 0.4$. 
The isolated leptons are considered to be separated from the jets and other leptons by $\Delta R_{\ell i} \gtrsim 0.4, i =j,\ell$.
For the background processes with
hard jets, proper MLM matching scheme \cite{Hoche:2006ph} has been chosen.
 \begin{table}
\centering
 \begin{tabular}{|c|c|c|c|c|c|}
  \hline
  & $\sin\alpha$ &  $v_t$ &   $m_3$ &  $m_5$  & $\sigma_{prod}$\\
  &  &  (in GeV) & (in GeV)  & (in GeV) & (in fb) \\
  \hline
  BP1 & 0.6 & 35 & 250 & 300 & 14.80\\ \hline
  BP2 & 0.5 & 45 & 300 & 350 & 18.53\\ 
  \hline
 \end{tabular}
	\caption{\it Benchmark points valid by all constraints and the corresponding production cross-section of our signal. }
	\label{tab:bp}
 \end{table}

To generate our signal and background events, we employ the following pre-selection cuts.
\begin{eqnarray}
 p_T(j,b) &>& 30 ~{\rm GeV}\,; \quad |\eta(j)| < 4.7 \,; \quad |\eta(b)| < 2.5 \,, \nonumber \\ 
 p_T(\ell) &>& 10 ~{\rm  GeV}\,, \quad |\eta(\ell)| <2.5 \,.
 \label{basic_cut}
\end{eqnarray}
The $b$-jets are tagged with the $p_T$-dependent $b$-tag efficiency following the medium criteria of Ref.~\cite{Sirunyan:2017ezt}
which has an average 75\% tagging efficiency of the $b$-jets with $50~{\rm GeV} < p_T < 200~{\rm GeV}$  and 1\% mis-tagging 
efficiency for light jets.

Additionally, we propose the following selection cuts to dis-entangle the signal from the SM backgrounds that would
enhance the signal significance. 

\begin{figure}[htbp!]
	\centering
\begin{subfigure}{.45\textwidth}
  \centering
  \includegraphics[width=0.95\linewidth]{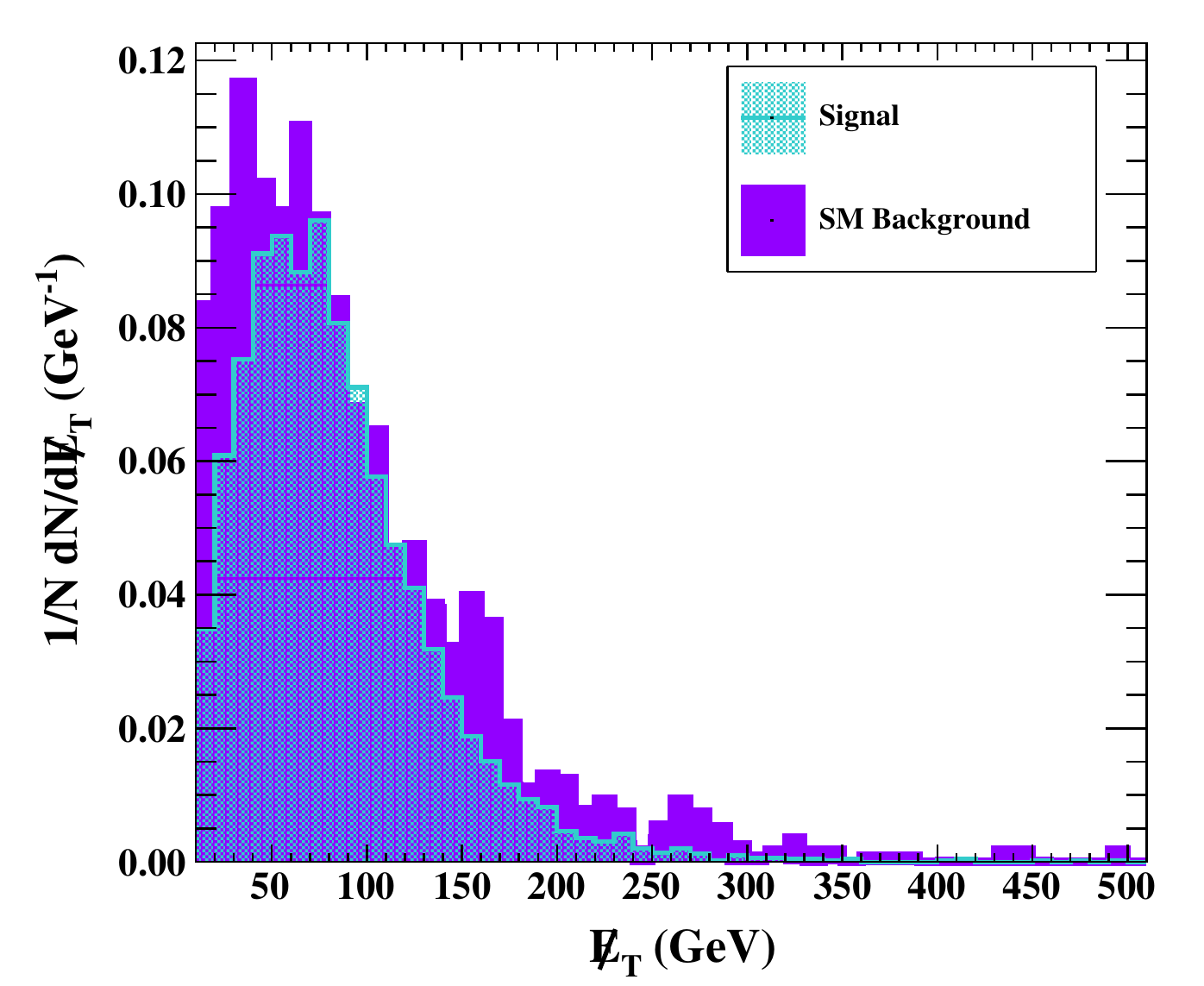}
  \caption{}\label{fig:met} 
		\end{subfigure}
\begin{subfigure}{.45\textwidth}
  \centering
  \includegraphics[width=0.95\linewidth]{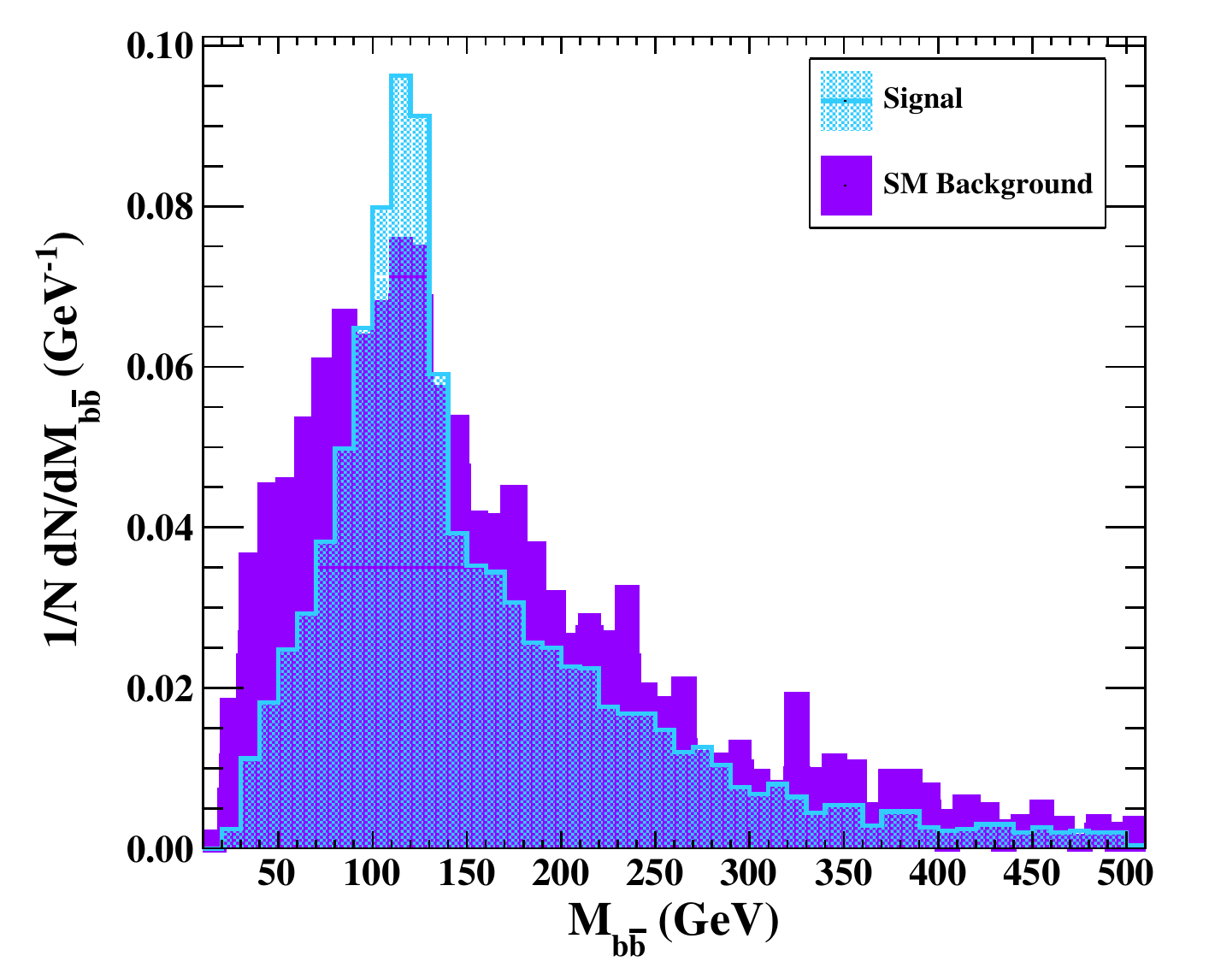}
  \caption{}\label{fig:invmass} 
  
  \end{subfigure}
	\caption{\textit{ Normalized distribution of the (left panel) Missing transverse energy $(\mET)$  and
			(right panel) the invariant mass of $b\bar{b}$ pair
			after the basic kinematical acceptance cuts(Eq.~\ref{basic_cut}) for the benchmark BP1.}}
	\label{fig:dist} 
\end{figure}

\begin{itemize}
           \item {\bf Selecting jets and leptons:}To affirm that our signal has 3 $b$ jets, we select 3 $b$ jets with $p_T(b) >$ 30 GeV 
           while rejecting any fourth $b$ jet with $p_T >$ 35 GeV. In addition to this, we demand two opposite sign leptons with 
           $p_T(\ell) >$ 10 GeV.
           
           \item {\bf Selecting $\mET$:} Both for signal and backgrounds, the missing energy($\mET$) comes from the 
           decay of $W^{\pm}$ boson. From the left panel of Fig.~\ref{fig:dist}, one can see that the distributions for signal 
           and backgrounds are peaking around similar $\mET$ value. However, the SM background has a large tail of the distribution. 
           We, thus, impose an upper bound of $\mET < 175$ GeV as an optimum cut to enhance the 
           signal over background ratio.
           
           \item {\bf Higgs mass reconstruction:}  In our signal, two $b$-tagged jets among the three are produced by the decay of the 
           Higgs. Therefore, the invariant mass of the two $b$-jets are expected to peak around 125 GeV. On the other hand,
           only the $t\bar t h$ can yield similar peak structure for the $M_{b\bar b}$. The right panel of Fig.~\ref{fig:dist}
           depicts the nature of $M_{b\bar b}$ distribution for both the signal and background. It is to be mentioned that
           the pair of b-jets is selected after rejecting any third b-jet that has the minimum azimuthal angle $\phi$. This follows
           from the argument that the third $b$-jet is a remnant of direct top decay in our signal and therefore expected to
           lie closer to the beam direction than the other two. Following Fig.~\ref{fig:dist}, we select 
           $100 ~\rm GeV < M_{b\bar{b}} < 150 ~\rm GeV$ for significant background reduction. This cut turns out to be 
           the most effective to reduce the large $t\bar t j$ background events. 
          \end{itemize}

The cut-flow is presented in Table.~\ref{tab:signal} for our two chosen benchmarks (BP1, BP2) of Table.~\ref{tab:bp}. The signal significance is 
defined as $\mathcal{S} = \frac{S}{\sqrt{S + B}}$, $S$ and $B$ being the number of signal and background events
after all the cuts respectively.

As is evident from Table.~\ref{tab:signal}, one can reach more than $5\sigma$ significance at the High Luminosity run of LHC (HL-LHC)
with 3000 $fb^{-1}$ luminosity and more than $3\sigma$ significance with only 1000 $fb^{-1}$  in the
specified parameter space of this model with this particular signal channel.
 \begin{table}[ht!]
	\centering
	\footnotesize
	\begin{tabular}{|p{3cm}|c|c|c|p{1.7cm}|}
		\cline{2-5}
		\multicolumn{1}{c|}{}& \multicolumn{3}{|c|}{Effective cross-section after the cut(fb)} & Significance reach at \\ \cline{1-4}
		SM-background  
		& Preselection cuts &$\mET $ cut  & Invmass cut   &  3(1) $ab^{-1}$  
		\\ \cline{1-4} 
		
		$t\bar{t}+$jet &124.41  & 119.52 &27.17 & integrated  \\ \cline{1-4} 
		$t\bar{t}h$ & 1.05 & 0.97 & 0.32  & luminosity \\ \cline{1-4} 
		$t\bar{t}W^{\pm}$ & 0.21  & 0.19 & 0.049  & at 14 TeV  \\ \cline{1-4} 
		$t\bar{t}Z$ & 0.84 & 0.79 & 0.21 & LHC  \\ \hline \hline
		${\rm Total~SM~Background}$ & 126.51 & 121.49 & 27.75 & --\\ \cline{1-5} \hline
		\multicolumn{1}{|c|}{$M_{H_{3^{\pm}}}=250 ~\rm GeV$}&  1.49  & 1.42 & 0.52 & 5.37 (3.10)  \\ \hline
		\multicolumn{1}{|c|}{$M_{H_{3^{\pm}}}=300 ~\rm GeV$}&  1.87  & 1.78 & 0.65 & 6.68 (3.86) \\ \hline 
		
	\end{tabular}
	\caption{\it No. of events obtained after each cut for both signal ($\ell^+ \ell^- + 3b + \mET $) and background and the significance obtained for $3  (1)~ ab^{-1}$ i$  $ntegrated luminosity at $\rm 14~TeV$ LHC.}
	\label{tab:signal}
\end{table}

 It is worth mentioning here that we did not consider the systematic uncertainty in background prediction while
calculating the signal significance. At the LHC Run-II, the current systematic uncertainty on our dominant SM background process $t\bar t + \rm {jet}$
can be considered to be around 10\% \cite{ATLAS:2018doi}. Inclusion of this uncertainty appropriately\cite{timmurphy.org} yield the 
signal significance for our becnhmark points for $M_{H_{3^{\pm}}}=250(300) ~\rm GeV$ respectively as 5.16(2.98)$\sigma$ and 6.46(3.73)$\sigma$
at 3(1)$\rm {ab^{-1}}$ luminosity. Therefore the inclusion of uncertainty will only make $\sim$4\% deviation from our quoted significance reach given in Table.~\ref{tab:signal}.
In fact, even within a conservative approximation of 20\% systematic uncertainty, the significance reach for our first signal point remains around $\approx 5\sigma$.
Moreover, it should also be noted that the actual systematic uncertainty
with the future high luminosity run of LHC at $\rm 14~TeV$ is currently unknown. 
Considering the future systematic uncertainty to be much lower than the current value \cite{PaganGriso:2642427}, 
our prediction on signal significance will not differ much from the realistic result and is useful for an preliminary understanding.

\section{Results}\label{sec:results}
\begin{figure}[ht!]
	\centering
	\includegraphics[width=8cm,height=6cm]{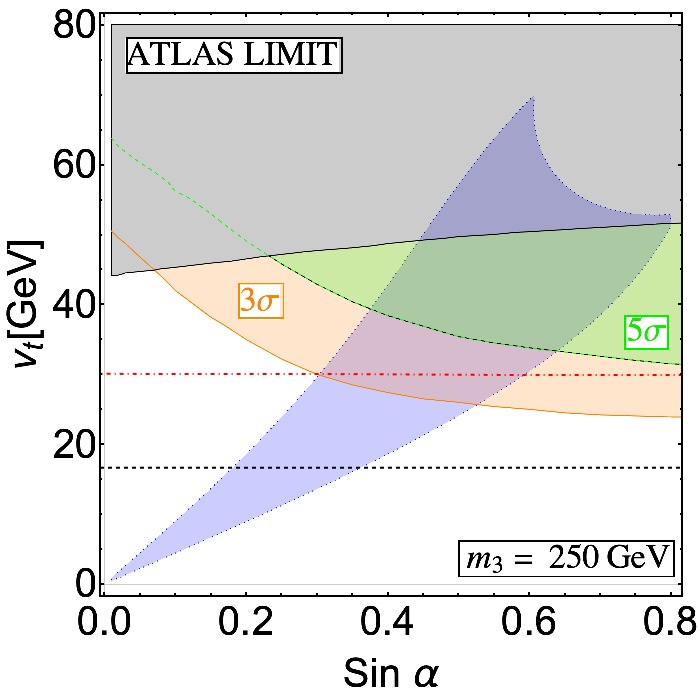}  ~
	\includegraphics[width=8cm,height=6cm]{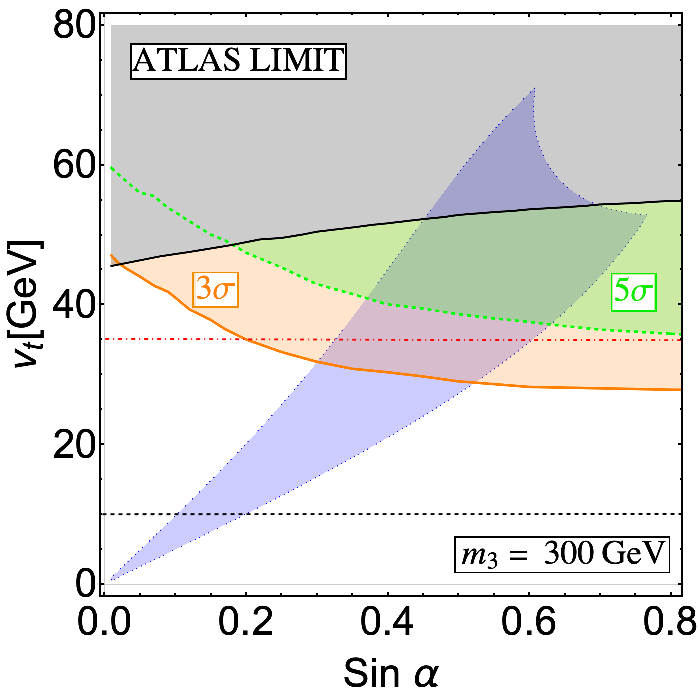} 
	\caption{\em The combined results on the model parameter space in the $(\sin\alpha-v_t)$ plane for charged Higgs masses $m_3 = 250$ GeV (left panel) and
		$m_3 = 300$ GeV (right panel). The color coding for Higgs data and theoretical constraint is
	same as in Fig.~\ref{f:hdata}. The gray shaded region with black solid boundary corresponds to the current ATLAS limit from charged Higgs analysis~\cite{Aaboud:2018cwk}. 
The Green(dashed) and Orange(Solid) lines are respectively the $5\sigma$ and $3\sigma$ signal significance reach at 14 TeV LHC with 3 $ab^{-1}$ luminosity
in the charged Higgs decaying to $Wh$ mode.}
	\label{f:combined}
\end{figure}

Let us now present our findings in a comprehensive manner. 
In Fig.~\ref{f:combined}, we show our complete results spanned in the model parameters $\{\sin\alpha-v_t\}$ plane for two particular choices of
$m_3$ and $m_5$ masses corresponding to our benchmark points in Table~\ref{tab:bp}. The region allowed by the latest LHC Higgs data~\cite{CMS:2018lkl,ATLAS:2018doi} and the
bounds from the theoretical constraints for the two distinct cases of trilinear terms follows the same color combinations as in Fig.~\ref{f:hdata}.
On top of it, we show the bounds imposed by the ATLAS analysis for 13 TeV data on charged Higgs searches in $tb$ channel~\cite{Aaboud:2018cwk} which is shown by the
gray shaded region bounded by the solid black line. 
As can be seen from the Fig.~\ref{f:hdata}, this bound can only restrict a small parameter space in association with the Higgs data. Quantitatively, 
for charged Higgs masses around 250 GeV, $v_t > 50 ~\rm GeV$ can only be discarded. For such $v_t$, the bound on $\sin\alpha$ though comes solely from the Higgs data. 
As already mentioned, for smaller $v_t$, Higgs data only allows large $\sin\alpha$ which can only be probed by the other significant decay mode of the charged scalar to
$Wh$ channel. Hence, we present the $3\sigma$ and $5\sigma$ discovery reach at the HL-LHC with 
3 $ab^{-1}$ luminosity from our proposed signal channel for charged Higgs decaying to $W^{\pm}~h$ mode, depicted as solid (orange) and dashed (green)
lines respectively. The areas above those lines shaded in Orange and Green describe the parameter
space that can be probed at the future HL-LHC with more than $3\sigma$ and $5\sigma$ signal significance respectively. To illustrate further, one can see that a
5$\sigma$ discovery is possible for $v_t \gtrsim $ 31(35) GeV for $m_3 =$250(300) GeV with 3 $ab^{-1}$ integrated luminosity. 
In a nutshell, Fig.~\ref{f:combined}
shows the complete allowed parameter space of GM model from theoretical constraints, latest LHC Higgs data and charged Higgs searches altogether, while 
simultaneously showing the future reach of charged Higgs searches for the unbounded parameter space. 

Before concluding, we should remark on the limit imposed by the CMS doubly charged Higgs search analysis at 13 TeV LHC with 35 $\rm fb^{-1}$ luminosity \cite{Sirunyan:2017ret}.
The 95\% C.L. upper limit has been given on the GM model based on the searches made for Vector Boson fusion production of a doubly charged Higgs decaying into a pair of same-sign W boson.
For a 200-300 $\rm GeV$ doubly charged Higgs, the limit is translated as an exclusion of parameter space above $v_t > 30$ GeV at 95\% C.L. This already is quite in tension with the model scenario
with vanishing trilinear terms ($M_1=M_2=0$) where the theoretical bounds only allow $v_t >30$ GeV.
Therefore, following our prediction on charged Higgs decaying to $W h$ mode in the future high luminosity run,  a non-observance of such a signal would definitely seal the fate of the model scenario for $M_1=M_2=0$.

\section{Summary }\label{sec:summary}
To summarise, we consider the GM model as a preferred variant of scalar triplet extension of the SM that preserves the custodial symmetry allowing for a larger
triplet vev unlike the usual HTMs. Such large triplet vev leads to interesting search prospects for the singly charged scalar of the model that can decay to SM fermion
with large coupling. The latest LHC search result on charged Higgs decaying to $tb$ channel thus impose an upper bound on the triplet vev. On the other hand,
the theoretical constraints on the model parameter restricts the triplet vev from the lower end and can be as strong as 30 GeV for vanishing triliinear term of the 
GM potential. Moreover, the theoretical constraints entails an upper bound on the triplet and quintuplet masses which puts significant upper bound on 
the singly and doubly charged Higgs mass. Additionally, the non-standard triplet scalars mixing will modify the SM-like Higgs coupling to the SM fermions and 
gauge bosons. Therefore, it is mandatory to match the model prediction for the SM-like Higgs coupling with the latest LHC Higgs data.  
In the first part of this paper, we show the complete limit on the model parameter space from both the theoretical constraints as well as the 13 TeV LHC Higgs results.
We show that these two constraints together put a stringent bound on the model parameter space in neutral mixing angle $\sin\alpha$ and triplet vev $v_t$ plane. 
 As shown, a large positive $\sin\alpha \gtrsim 0.1(0.3)$ with $v_t>10 (30)~\rm GeV$ is preferred by these bounds for the two distinct cases of vanishing and non-vanishing trilinear terms. 
 In the second part of our paper, we intended to combine these limits with the latest bound
 from LHC on the charged scalar searches to $tb$ channel. We show that this limit excludes $v_t > 50$ GeV for $\sin\alpha \lesssim 0.4$ for a  250 GeV charged Higgs
 in the region allowed by the theoretical constraints and the Higgs data. 
 We also described that for a larger $\sin\alpha$, the singly charged Higgs predominantly decays to $Wh$ channel.
 Therefore, we showed a simple cut-based signal-to-background analysis for the singly charged Higgs decaying to $Wh$ mode. Our analysis revealed that this search channel can probe
 the unbounded region of the model parameter by all the above-mentioned constraints at $5\sigma$ signal significance at the future high luminosity run of the LHC. 
\section*{Acknowledgements}
IS would like to thank Michihisa Takeuchi and Satoshi Shirai for useful discussions. 
The work of IS was supported by World Premier International Research Center Initiative (WPI), MEXT, Japan. 
NG would like to acknowledge the Council of Scientific and Industrial Research (CSIR), Government of India 
for financial support.   
SG would like to thank the University Grants Commission, Government of India, for a research fellowship.



\providecommand{\href}[2]{#2}\begingroup\raggedright\endgroup

\end{document}